\documentclass[sigconf]{acmart}

\usepackage{enumitem}

\usepackage{amssymb}

\usepackage{tabularx} 
\usepackage{lipsum}
\usepackage{pifont}
\usepackage{url}
\usepackage{multirow}
\usepackage{subfigure}
\usepackage{bm}
\usepackage{verbatim}
\usepackage{ragged2e}
\usepackage{caption}
\usepackage{subcaption}
\usepackage{graphicx}
\usepackage[normalem]{ulem}
\useunder{\uline}{\ul}{}
\usepackage{amsmath}
\usepackage{longtable}
\usepackage{adjustbox}
\usepackage{microtype}
\usepackage{setspace}
\usepackage{float}
\usepackage{balance}
\usepackage{algorithm}
\usepackage{algpseudocode}
\usepackage{booktabs}
\usepackage{multirow}
\usepackage{hyperref}

\newcommand{\aka}

\newcommand{\bym}[1]{\textcolor{black}{#1}}
\newcommand{\csy}[1]{\textcolor{black}{#1}}

\AtBeginDocument{%
  \providecommand\BibTeX{{%
    \normalfont B\kern-0.5em{\scshape i\kern-0.25em b}\kern-0.8em\TeX}}}

\copyrightyear{2018}
\setcopyright{acmlicensed}
\acmYear{2018}
\acmDOI{XXXXXXX.XXXXXXX}
\acmConference[Conference acronym 'XX]{Make sure to enter the correct
  conference title from your rights confirmation emai}{June 03--05,
  2018}{Woodstock, NY}
\acmBooktitle{Woodstock '18: ACM Symposium on Neural Gaze Detection,
 June 03--05, 2018, Woodstock, NY} 
\acmISBN{978-1-4503-XXXX-X/18/06}

\begin{document}

\title{Unveiling Inference Scaling for Difference-Aware User Modeling in LLM Personalization}

\author{Suyu Chen}
\orcid{0009-0006-8173-3717}
\affiliation{
  \institution{University of Science and Technology of China}
  \city{Hefei}
  \country{China}
}
\email{chensuyv@mail.ustc.edu.cn}

\author{Yimeng Bai}
\orcid{0009-0008-8874-9409}
\affiliation{
  \institution{University of Science and Technology of China}
  \city{Hefei}
  \country{China}
}
\email{baiyimeng@mail.ustc.edu.cn}
\authornote{Equal contribution.}

\author{Yulong Huang}
\orcid{0009-0007-8835-8558}
\affiliation{
  \institution{University of Science and Technology of China}
  \city{Hefei}
  \country{China}
}
\email{huangyulong@mail.ustc.edu.cn}

\author{Xiaoyan Zhao$^\dag$}
\orcid{0000-0001-6001-1260}
\affiliation{
  \institution{The Chinese University of Hong Kong}
  \city{Hong Kong}
  \country{China}}
\email{xzhao@se.cuhk.edu.hk}

\author{Yang Zhang}
\orcid{0000-0002-7863-5183}
\affiliation{
  \institution{National University of Singapore}
  \city{Singapore}
  \country{Singapore}
}
\email{zyang1580@gmail.com}
\authornote{Corresponding author.}

\def\authors{Suyu Chen, Yimeng Bai, Yulong Huang, Xiaoyan Zhao, and Yang Zhang}


\renewcommand{\shortauthors}{Suyu Chen et al.}

\begin{abstract}

Large Language Models (LLMs) are increasingly integrated into users’ daily lives, driving a growing demand for personalized outputs. Prior work has primarily leveraged a user’s own history, often overlooking inter-user differences that are critical for effective personalization. While recent methods have attempted to model such differences, their feature extraction processes typically rely on fixed dimensions and quick, intuitive inference (System-1 thinking), limiting both the coverage and granularity of captured user differences. To address these limitations, we propose \textit{Difference-aware Reasoning Personalization} (DRP), a framework that reconstructs the difference extraction mechanism by leveraging inference scaling to enhance LLM personalization. DRP autonomously identifies relevant difference feature dimensions and generates structured definitions and descriptions, enabling slow, deliberate reasoning (System-2 thinking) over user differences. Experiments on personalized review generation demonstrate that DRP consistently outperforms baseline methods across multiple metrics. 

\end{abstract}





\maketitle

\section{Introduction}

Large Language Models (LLMs) have achieved remarkable success across real-world applications---such as virtual assistants~\cite{seed1.5}, content creation~\cite{seedcoder}, and scientific discovery~\cite{deepreport}---driven by their powerful capabilities in context comprehension, generation, and reasoning~\cite{qwen3,deepseekr1}. However, existing LLMs are generally built following a ``one-size-fits-all'' paradigm~\cite{pllmsurvey}, failing to account for the variability in user preferences. As the demand for delivering tailored and engaging experiences grows, interest in \textit{LLM personalization} has steadily increased across both academic and industrial domains~\cite{alignx,lamp}, aiming to tailor model outputs based on user-specific information. 

Most existing works on LLM personalization adopt the \textit{memory-retrieval} paradigm~\cite{intsum,personadb}, where user history is stored in memory, and key information is then retrieved  as instructional contexts to customize model responses. Early approaches focused exclusively on leveraging information about the individual user for personalization~\cite{teach,pearl}. More recent works, such as DPL~\cite{dpl}, emphasize that effective personalization requires identifying the differences between users that shape individual preferences, an insight supported by research in psychology and behavioral science~\cite{uniq}. Accordingly, DPL selects representative users for comparison, extracts task-relevant differences, and incorporates them as augmented features within prompts to improve personalization. 

Despite demonstrated effectiveness of DPL, we argue that its difference extraction mechanism remains inherently limited in both the \textit{coverage} and \textit{granularity} of generated features. In terms of coverage, the extracted differences are restricted to fixed, pre-defined feature dimensions (writing, emotional, and semantic style), making it challenging to capture the potentially 
\csy{unbounded} space of user preferences~\cite{alignx}. Regarding granularity, the mechanism relies on quick inference learned at training time (also referred to as System-1 thinking)~\cite{thinkingfastslow}, and therefore lacks the capability to deeply analyze fine-grained patterns that may be critical, such as the underlying causes driving the observed differences~\cite{inferrec}. Consequently, in dynamic personalization scenarios, it is desirable to leverage the advanced reasoning capabilities at inference time (System-2 thinking), enabling on-the-fly reasoning to uncover richer and more informative difference features.

To address these limitations, we propose reconstructing DPL’s difference extraction mechanism by incorporating additional test-time computation to enable flexible feature dimensions and extended reasoning chains. This approach is inspired by the success of \textit{inference scaling} in math~\cite{openr} and coding~\cite{o1-coder} tasks. First, by enabling flexible dimension generation, the model can more thoroughly explore the user preference space, alleviating the constraints imposed by handcrafted dimensions. Second, by allowing sophisticated reasoning, the model can produce fine-grained rationales, offering deeper insights into user differences. Together, this opens up the possibility of scaling both the coverage and granularity of generated difference features in a training-free manner, thereby capturing a broader spectrum of user preferences.

To this end, we propose \textit{Difference-aware Reasoning Personalization} (DRP), a novel framework that enhances LLM personalization through inference scaling. Specifically, DRP leverages \bym{reasoning-enhanced} LLMs as the difference extractor: it first autonomously identifies the appropriate feature dimensions for generating differences, and then provides clear definitions and descriptions for each dimension. This design provides a natural entry point for the model to transition from quick, intuitive inference (System-1 thinking) to slow, deliberate reasoning (System-2 thinking), thereby enabling scalable personalization. We compare DRP’s performance using reasoning-enhanced models (DeepSeek-R1-Distill-Qwen~\cite{deepseekr1}) versus standard instruction-tuned models (Qwen2.5-Instruct~\cite{qwen2.5}) across parameter sizes from 1.5B to 32B, on the representative personalization task of review generation, following prior studies~\cite{dpl,latentdpl,nextquill}. Across these settings, DRP consistently outperforms baselines, with reasoning-enhanced LLMs achieving gains of up to 23.0\% on key metrics such as BLEU, and quantitative analyses provide empirical validation that these improvements stem from both broader coverage and enhanced granularity of user difference features.

The main contributions of this work are summarized as follows:

\begin{itemize}[leftmargin=*]

\item To the best of our knowledge, we are the first to explore inference scaling to overcome limitations of existing LLM personalization methods, including restricted coverage and limited granularity in capturing user differences.

\item We propose DRP, which autonomously determine difference feature dimensions and generate structured definitions and descriptions through extended reasoning, facilitating scalable difference features for personalization.

\item We conduct a series of experiments, showcasing the effectiveness
of DRP in LLM personalization.

\end{itemize}
\begin{figure}[t]
\centering
\includegraphics[width=1.00\linewidth]{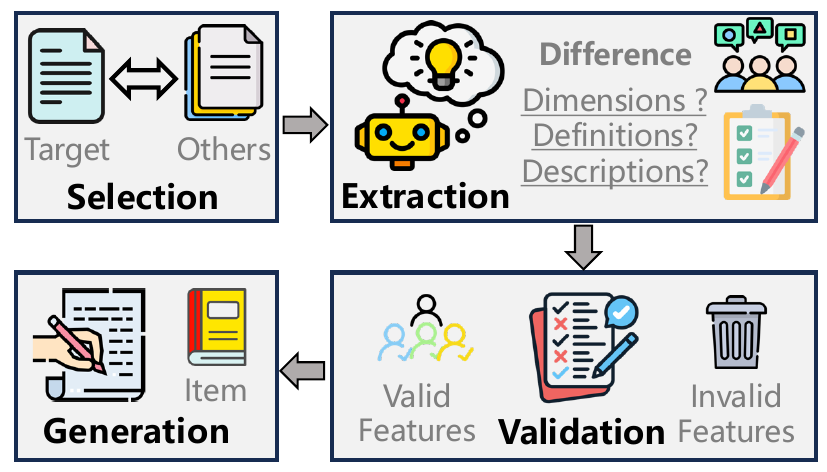}
\caption{Overview of our proposed DRP Framework, where the core innovation lies in automatic dimension discovery and reasoning-based difference extraction.} 
\label{fig:process}
\end{figure}
\section{Methodology}
 
\subsection{Task Formulation}
LLM personalization aims to generate content that reflects a user’s individual preferences. In this work, we focus on personalized review generation~\cite{dpl}. Each user has historical texts—either written or preferred by them—that capture their personal style and preferences. Let $\mathcal{D}$ denote the historical dataset with samples $(u, i, y)$, where $u$ is a user, $i$ an item, and $y$ the user-associated text. Given a target user $u$ and item $i$, the objective is to generate a personalized review $\hat{y}$ that aligns with the user’s historical texts. Following standard retrieval-based personalization~\cite{lamp}, the target user’s key historical data $\mathcal{D}_u^\star$ can be retrieved as contextual input, forming the base generation mechanism:
\begin{equation}
\hat{y} = \text{LLM}_G(u, i, \mathcal{D}_u^\star),
\end{equation}
which serves as the foundation upon which DRP builds.

\subsection{DRP Framework}
We propose DRP, a framework for difference-aware user modeling that enhances personalization via automatic dimension discovery and reasoning-based feature extraction.

\textbf{Step 0: Representative User Selection.}  
To extract meaningful differences, DRP selects multiple users $\mathcal{R}$ for comparison. Users are clustered \bym{using K-means} based on their historical text embeddings, and for a target user $u$, $M$ representatives are chosen from all clusters excluding $u$’s own:
\begin{equation}
\mathcal{R} = \{r_1, r_2, \dots, r_M\}.
\end{equation}

\textbf{Step 1: Reasoning-based Difference Extraction.}  
Unlike prior work like DPL~\cite{dpl}, which relies on fixed, handcrafted feature dimensions and rapid System-1 inference, DRP combines automatic dimension discovery with deliberate reasoning in a single step. The reasoning-enhanced extractor $\text{LLM}_E^R$ identifies relevant difference dimensions $\Delta_\text{auto}$ and analyzes the underlying factors driving the differences between the target user and each representative:
\begin{equation}
\{\delta_{u,r} = \text{LLM}_E^R(\mathcal{D}_u^\star, \mathcal{D}_r^\star, \Delta_\text{auto}) \mid r \in \mathcal{R}\}.
\end{equation}
This produces structured definitions and descriptions, enabling scalable difference features for personalization.

\textbf{Step 2: Reflective Validation.}  
The extracted difference features are independently verified by a reflection-enabled LLM, which filters out invalid or spurious differences. Formally:
\begin{equation}
\{\tilde{\delta}_{u,r}\}_{r \in \mathcal{R}} = \text{LLM}_V(\{\delta_{u,r}\}_{r \in \mathcal{R}}),
\end{equation}
\bym{where $\text{LLM}_V$ denotes the LLM performing the validation (defaulting to the same model as $\text{LLM}_E$).}

\textbf{Step 3: Personalized Generation.}  
The validated difference features $\{\tilde\delta_{u,r}\}_{r \in \mathcal{R}}$ and the target user’s key history $\mathcal{D}_u^\star$ are jointly summarized by \csy{the summarizer} $\text{LLM}_S$ and used by the generator $\text{LLM}_G$ to produce the final personalized review:
\begin{equation}
\hat{y} = \text{LLM}_G\Big(u, i, \mathcal{D}_u^\star, \text{LLM}_S(\mathcal{D}_u^\star, \{\tilde\delta_{u,r}\}_{r \in \mathcal{R}})\Big).
\end{equation}

\begin{table*}[t]
\centering
\footnotesize 
\setlength{\tabcolsep}{3.0pt} 
\renewcommand{\arraystretch}{1.05} 
\caption{Results on both datasets. QwenX and DpSkX refer to the Qwen-Instruct and DeepSeek-R1-Distill-Qwen models, respectively, each with X parameters. The best and second-best results are highlighted in bold and underlined font, respectively.}
\label{tab:main_results}

\begin{tabular}{llccccccccccccccc}
\toprule
& & \multicolumn{7}{c}{\textbf{Baseline Methods}} & \multicolumn{8}{c}{\textbf{DRP (ours)}} \\
\cmidrule(lr){3-9}\cmidrule(lr){10-17}
\raisebox{1.9ex}[0pt]{\textbf{Dataset}} & \raisebox{1.9ex}[0pt]{\textbf{Metric}}
& Non-P & RAG & IntSum & LLM-TRSR & CICL & P-DB & DPL 
& Qwen1.5B & Qwen7B & Qwen14B & Qwen32B & DpSk1.5B & DpSk7B & DpSk14B & DpSk32B \\
\midrule

\multirow{4}{*}{\textbf{Books}}
& BLEU      & 1.2616 & 4.1082 & 4.6256 & 4.6268 & 5.1555 & 5.1862 & 5.6534
            & 5.2209 & 6.0444 & 6.1955 & \underline{6.4390} & 4.8659 & 5.2365 & 6.2846 & \textbf{6.9535} \\
& METEOR    & 0.1511 & 0.2171 & 0.2280 & 0.2250 & 0.2339 & 0.2368 & 0.2373
            & 0.2303 & 0.2457 & 0.2435 & \underline{0.2469} & 0.2269 & 0.2349 & 0.2449 & \textbf{0.2551} \\
& ROUGE-1   & 0.2711 & 0.3238 & 0.3183 & 0.3140 & 0.3262 & 0.3271 & 0.3289
            & 0.3274 & 0.3387 & 0.3386 & \underline{0.3394} & 0.3192 & 0.3244 & 0.3361 & \textbf{0.3424} \\
& ROUGE-L   & 0.1457 & 0.1636 & 0.1614 & 0.1603 & 0.1664 & 0.1672 & 0.1700
            & 0.1689 & 0.1728 & 0.1768 & \underline{0.1787} & 0.1645 & 0.1682 & 0.1777 & \textbf{0.1818} \\
\midrule

\multirow{4}{*}{\shortstack{\textbf{CDs \&}\\\textbf{Vinyl}}}
& BLEU      & 0.4783 & 1.9271 & 2.3621 & 2.3628 & 2.2978 & 2.1616 & 2.7189
            & 2.2435 & 2.7000 & 2.7561 & 2.8014 & 2.2621 & 2.5999 & \textbf{2.8617} & \underline{2.8092} \\
& METEOR      & 0.1327 & 0.1857 & 0.1956 & 0.1952 & 0.1946 & 0.1912 & 0.2026
            & 0.1919 & 0.2044 & 0.2059 & \textbf{0.2080} & 0.1936 & 0.2009 & 0.2062 & \underline{0.2073} \\
& ROUGE-1       & 0.2396 & 0.2914 & 0.2958 & 0.2937 & 0.2986 & 0.2935 & 0.3019
            & 0.2989 & 0.3105 & \textbf{0.3120} & \underline{0.3119} & 0.2973 & 0.3029 & 0.3112 & 0.3111 \\
& ROUGE-L       & 0.1273 & 0.1418 & 0.1422 & 0.1417 & 0.1433 & 0.1421 & 0.1453
            & 0.1450 & 0.1473 & \underline{0.1483} & 0.1480 & 0.1438 & 0.1453 & 0.1480 & \textbf{0.1515} \\
\bottomrule
\end{tabular}
\end{table*}

\section{Experiment}

In this section, we conduct experiments to answer the following research questions:

\noindent \textbf{RQ1}: Does DRP outperform baseline methods in personalized text generation when leveraging inference scaling?

\noindent \textbf{RQ2}: Does DRP improve the coverage and granularity of difference extraction to enhance personalization performance?

\subsection{Experimental Setup}

\noindent
\textbf{Datasets.} 
Building upon prior work~\cite{lamp}, we focus on the representative task of item review generation for LLM personalization. Specifically, we utilize the Amazon Reviews dataset~\cite{amazon2023} preprocessed by DPL~\cite{dpl}, which includes the \textit{Books} and \textit{CDs \& Vinyl}.

\noindent
\textbf{Baselines.}
We compare DRP with the following methods: (1) Non-personalized approach: \textit{Non-P} ; (2) RAG-based variants: \textit{RAG}~\cite{lamp}, \textit{IntSum}~\cite{intsum}, \textit{LLM-TRSR}~\cite{llm-trsr}, \textit{CICL}~\cite{cicl}, \textit{P-DB}~\cite{personadb} ; (3) Difference-aware method: \textit{DPL}~\cite{dpl}.

\noindent
\textbf{Evaluation Metrics.} 
Following previous works on personalized text generation~\cite{dpl,latentdpl}, we evaluate all methods using \textit{BLEU, METEOR, ROUGE-1, ROUGE-L}.

\noindent
\textbf{Implementation Details}.
In DRP, the difference extraction is performed using reasoning-enhanced (DeepSeek-R1-Distill-Qwen~\cite{deepseekr1}) and standard (Qwen2.5-Instruct~\cite{qwen2.5}) models across 1.5B to 32B parameter scales. For each model, experiments are conducted at temperatures 0 and 0.8, with results averaged. The remaining configurations are consistent with DPL. Due to space limitations, details regarding the datasets, baselines, and prompts used in the experiments are provided in the repository\footnote{\url{https://github.com/Chen-Suyu/DRP}}.

\subsection{Main Result (RQ1)}

Comparison results of the overall performance are presented in Table~\ref{tab:main_results}, from which the following observations can be drawn: 

\textbf{Superior Performance}. 
DRP consistently outperforms all baselines on both datasets after scaling to 14B parameters (the default in DPL),
demonstrating that the inference scaling of difference extractor significantly improves personalized generation quality, with a maximum BLEU gain of \textbf{23.0\%} on Books.

\textbf{Scaling and Reasoning Effects}. DRP’s performance improves with larger model sizes, reflecting the benefits of increased capacity. At the same extractor size, 
DeepSeek outperforms Qwen \csy{in most cases}, likely due to its reasoning-enhanced design, which enables finer-grained extraction of user-specific differences and richer contextual information for personalized generation.

\begin{figure}[t]
    \centering
    \includegraphics[height=0.44\linewidth,width=1.00\linewidth]{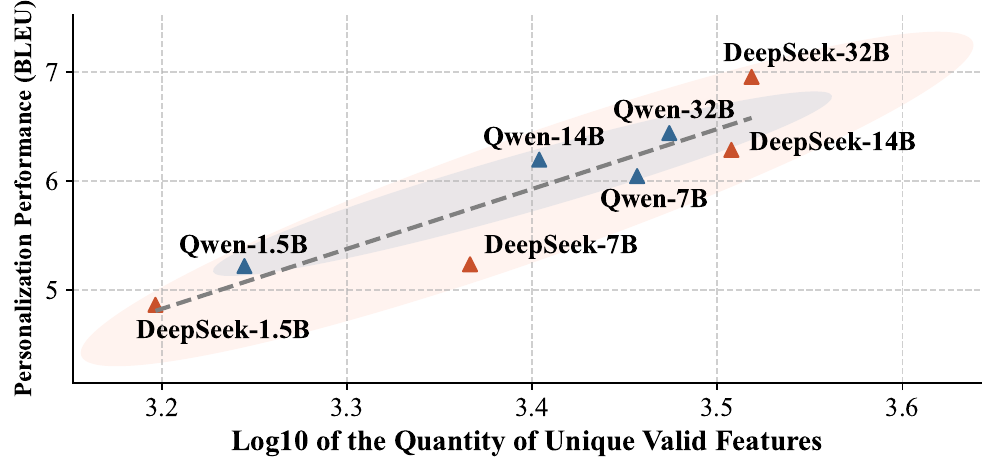}
    \caption{Relationship between BLEU and UVQ across DRP backbones on Books.}
    \label{fig:valid_features}
\end{figure}
\begin{figure}[t]
    \centering
    \includegraphics[height=0.4\linewidth,width=1.0\linewidth]{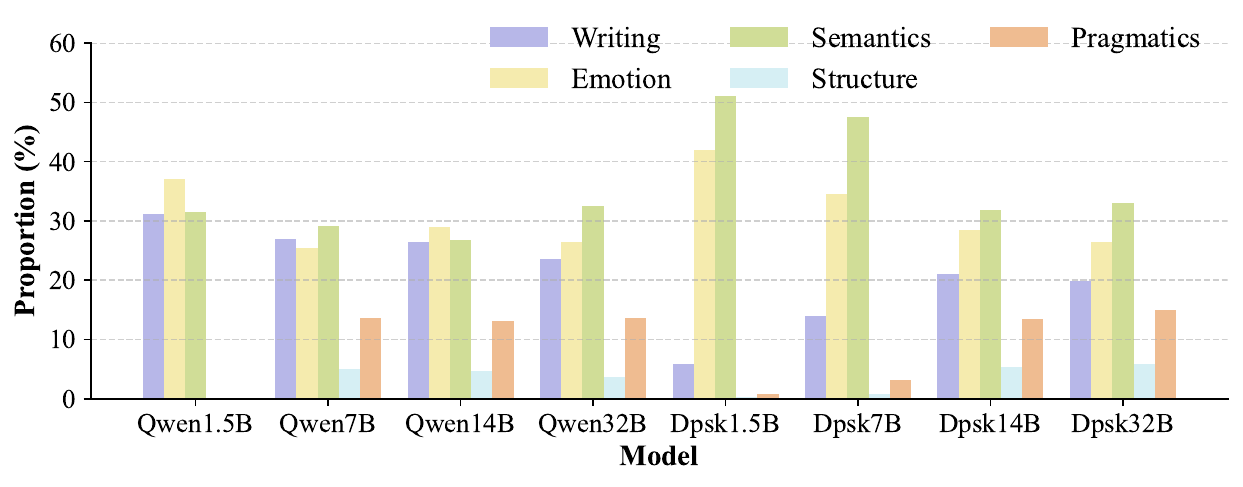}
    \caption{Proportions of difference feature categories across DRP backbones on Books.}
    \label{fig:five}
    \vspace{-8pt}
\end{figure}

\subsection{Qualitative Analysis (RQ2)}

\subsubsection{Coverage Analysis}

We propose a metric to quantify the coverage of generated difference features. Based on a posterior analysis, we categorize the features into five representative types: \textit{Writing}, \textit{Emotion}, \textit{Semantics}, \textit{Structure}, and \textit{Pragmatics}. The first two capture surface-level tendencies, while the latter three reflect deeper cognitive and communicative patterns.

A difference feature is considered \textit{valid} if it is \textit{comparative} (reflecting the target relative to others), \textit{atomic} (capturing a single property), \textit{clear} (unambiguous), \textit{categorized} (belonging to one of the five predefined types), and \textit{consistent} (directionally stable across multiple sources). For each user, we deduplicate valid features and remove conflicting directional terms. Aggregating the remaining unique valid features across the dataset gives the \textit{Unique Valid Quantity} (UVQ). A higher UVQ indicates that the model captures a broader and more diverse set of difference features, reflecting better coverage of potential user differences.

We evaluate these features using Qwen-2.5-Instruct-32B via in-context learning. Figure~\ref{fig:valid_features} shows the relationship between BLEU and UVQ across different extractor models on the Books dataset. A significant positive correlation can be observed between BLEU and UVQ, indicating that one source of improvement in our proposed method stems from an increased number of unique valid differential features, which validates the necessity of overcoming the limitations imposed by handcrafted feature dimensions.

\subsubsection{Granularity Analysis}

Then, we analyze the granularity of the difference features. We report the proportions of the five dimension categories described above across different extractor models, as shown in Figure~\ref{fig:five}. It is evident that the stronger DeepSeek variants exhibit higher proportions in Semantics, Structure, and Pragmatics, suggesting that reasoning allows the model to capture deeper cognitive patterns during thinking processing, rather than merely surface-level tendencies. Additionally, we present a case study in Table~\ref{tab:drp-case}, which demonstrates that the reasoning process can generate analyses of the underlying causes driving user differences, thereby uncovering rare but valuable personalized features.

\begin{table}[t]
\caption{Case study of difference features generated by DRP. }
\centering
\scriptsize
\setlength{\tabcolsep}{3pt}
\renewcommand{\arraystretch}{1.05}

\begin{tabularx}{\columnwidth}{@{}>{\raggedright\arraybackslash}X >{\raggedright\arraybackslash}X@{}}
\toprule
\multicolumn{1}{c}{\textbf{Qwen2.5-14B-Instruct}} &
\multicolumn{1}{c}{\textbf{DeepSeek-R1-Distill-Qwen-14B}} \\
\midrule

“…includes more contextual information about their personal experiences (e.g., podcasts about true crime)… Others provide a more direct and concise summary…”
&
“…provides more background context (podcasts/other books), \textbf{leading to higher information density}… Others are more concise, focusing on plot/characters.” \\
\cmidrule(lr){1-2}

“…notably more enthusiastic and positive (‘rave reviews’, ‘really cool books’)… Others are more measured and sometimes critical.”
&
“…highly enthusiastic (exclamation marks + strong positive adjectives)… Others more measured, some neutral or slightly negative \textbf{due to specific criticisms}.” \\

\bottomrule
\end{tabularx}
\label{tab:drp-case}
\end{table}
\section{Conclusion}

We proposed DRP to enhance LLM personalization by leveraging inference scaling. DRP autonomously identified flexible difference dimensions and generated structured definitions, enabling the transition from quick, intuitive inference (System-1 thinking) to slow, deliberate reasoning (System-2 thinking). Overall, DRP demonstrated the potential to scale both the coverage and granularity of user difference features, facilitating scalable personalization.

\bibliographystyle{ACM-Reference-Format}
\balance
\bibliography{8_reference}

@String{Computing="Computing" }

@String{Macmillan="Macmillan" }

@article{seed1.5,
title={Seed1. 5-vl technical report},
author={Guo, Dong and Wu, Faming and Zhu, Feida and Leng, Fuxing and Shi, Guang and Chen, Haobin and Fan, Haoqi and Wang, Jian and Jiang, Jianyu and Wang, Jiawei and others},
journal={arXiv preprint arXiv:2505.07062},
year={2025}
}

@article{seedcoder,
title={Seed-coder: Let the code model curate data for itself},
author={Seed, ByteDance and Zhang, Yuyu and Su, Jing and Sun, Yifan and Xi, Chenguang and Xiao, Xia and Zheng, Shen and Zhang, Anxiang and Liu, Kaibo and Zan, Daoguang and others},
journal={arXiv preprint arXiv:2506.03524},
year={2025}
}

@article{qwen3,
title={Qwen3 technical report},
author={Yang, An and Li, Anfeng and Yang, Baosong and Zhang, Beichen and Hui, Binyuan and Zheng, Bo and Yu, Bowen and Gao, Chang and Huang, Chengen and Lv, Chenxu and others},
journal={arXiv preprint arXiv:2505.09388},
year={2025}
}

@article{deepseekr1,
title={Deepseek-r1 incentivizes reasoning in llms through reinforcement learning},
author={Guo, Daya and Yang, Dejian and Zhang, Haowei and Song, Junxiao and Wang, Peiyi and Zhu, Qihao and Xu, Runxin and Zhang, Ruoyu and Ma, Shirong and Bi, Xiao and others},
journal={Nature},
volume={645},
number={8081},
pages={633--638},
year={2025},
publisher={Nature Publishing Group UK London}
}

@inproceedings{deepreport,
author = {Xu, Yi and Fu, Luoyi and Sheng, Shuqian and Xue, Bo and Ding, Jiaxin and Zhou, Lei and Wang, Xinbing and Zhou, Chenghu},
title = {DeepReport: An AI-assisted Idea Generation System for Scientific Research},
year = {2025},
isbn = {9798400715921},
publisher = {Association for Computing Machinery},
address = {New York, NY, USA},
booktitle = {Proceedings of the 48th International ACM SIGIR Conference on Research and Development in Information Retrieval},
pages = {3969–3973},
numpages = {5},
location = {Padua, Italy},
series = {SIGIR '25}
}

@article{pllmsurvey,
title={A survey of personalized large language models: Progress and future directions},
author={Liu, Jiahong and Qiu, Zexuan and Li, Zhongyang and Dai, Quanyu and Yu, Wenhao and Zhu, Jieming and Hu, Minda and Yang, Menglin and Chua, Tat-Seng and King, Irwin},
journal={arXiv preprint arXiv:2502.11528},
year={2025}
}

@article{alignx,
title={From 1,000,000 users to every user: Scaling up personalized preference for user-level alignment},
author={Li, Jia-Nan and Guan, Jian and Wu, Songhao and Wu, Wei and Yan, Rui},
journal={arXiv preprint arXiv:2503.15463},
year={2025}
}

@inproceedings{lamp,
title = "{L}a{MP}: When Large Language Models Meet Personalization",
author = "Salemi, Alireza  and
Mysore, Sheshera  and
Bendersky, Michael  and
Zamani, Hamed",
editor = "Ku, Lun-Wei  and
Martins, Andre  and
Srikumar, Vivek",
booktitle = "Proceedings of the 62nd Annual Meeting of the Association for Computational Linguistics (Volume 1: Long Papers)",
month = aug,
year = "2024",
address = "Bangkok, Thailand",
publisher = "Association for Computational Linguistics",
pages = "7370--7392",
}

@article{intsum,
title={Integrating summarization and retrieval for enhanced personalization via large language models},
author={Richardson, Chris and Zhang, Yao and Gillespie, Kellen and Kar, Sudipta and Singh, Arshdeep and Raeesy, Zeynab and Khan, Omar Zia and Sethy, Abhinav},
journal={arXiv preprint arXiv:2310.20081},
year={2023}
}

@inproceedings{personadb,
title = "Persona-{DB}: Efficient Large Language Model Personalization for Response Prediction with Collaborative Data Refinement",
author = "Sun, Chenkai  and
Yang, Ke  and
Gangi Reddy, Revanth  and
Fung, Yi  and
Chan, Hou Pong  and
Small, Kevin  and
Zhai, ChengXiang  and
Ji, Heng",
booktitle = "Proceedings of the 31st International Conference on Computational Linguistics",
month = jan,
year = "2025",
address = "Abu Dhabi, UAE",
publisher = "Association for Computational Linguistics",
pages = "281--296",
}

@article{teach,
title={Teach LLMs to Personalize--An Approach inspired by Writing Education},
author={Li, Cheng and Zhang, Mingyang and Mei, Qiaozhu and Wang, Yaqing and Hombaiah, Spurthi Amba and Liang, Yi and Bendersky, Michael},
journal={arXiv preprint arXiv:2308.07968},
year={2023}
}

@inproceedings{pearl,
title = "Pearl: Personalizing Large Language Model Writing Assistants with Generation-Calibrated Retrievers",
author = "Mysore, Sheshera  and
Lu, Zhuoran  and
Wan, Mengting  and
Yang, Longqi  and
Sarrafzadeh, Bahareh  and
Menezes, Steve  and
Baghaee, Tina  and
Gonzalez, Emmanuel Barajas  and
Neville, Jennifer  and
Safavi, Tara",
booktitle = "Proceedings of the 1st Workshop on Customizable NLP: Progress and Challenges in Customizing NLP for a Domain, Application, Group, or Individual (CustomNLP4U)",
month = nov,
year = "2024",
address = "Miami, Florida, USA",
publisher = "Association for Computational Linguistics",
pages = "198--219",
}

@inproceedings{dpl,
title = "Measuring What Makes You Unique: Difference-Aware User Modeling for Enhancing {LLM} Personalization",
author = "Qiu, Yilun  and
  Zhao, Xiaoyan  and
  Zhang, Yang  and
  Bai, Yimeng  and
  Wang, Wenjie  and
  Cheng, Hong  and
  Feng, Fuli  and
  Chua, Tat-Seng",
editor = "Che, Wanxiang  and
  Nabende, Joyce  and
  Shutova, Ekaterina  and
  Pilehvar, Mohammad Taher",
booktitle = "Findings of the Association for Computational Linguistics: ACL 2025",
month = jul,
year = "2025",
address = "Vienna, Austria",
publisher = "Association for Computational Linguistics",
pages = "21258--21277",
ISBN = "979-8-89176-256-5",
}

@article{uniq,
title={You like what I like, but I don’t like what you like: Uniqueness motivations in product preferences},
author={Irmak, Caglar and Vallen, Beth and Sen, Sankar},
journal={Journal of Consumer Research},
volume={37},
number={3},
pages={443--455},
year={2010},
publisher={The University of Chicago Press}
}

@book{thinkingfastslow,
title={Thinking, fast and slow},
author={Kahneman, Daniel},
year={2011},
publisher={macmillan}
}

@article{inferrec,
title={Inference computation scaling for feature augmentation in recommendation systems},
author={Liu, Weihao and Du, Zhaocheng and Zhao, Haiyuan and Zhang, Wenbo and Zhao, Xiaoyan and Wang, Gang and Dong, Zhenhua and Xu, Jun},
journal={arXiv preprint arXiv:2502.16040},
year={2025}
}

@article{openr,
title={Openr: An open source framework for advanced reasoning with large language models},
author={Wang, Jun and Fang, Meng and Wan, Ziyu and Wen, Muning and Zhu, Jiachen and Liu, Anjie and Gong, Ziqin and Song, Yan and Chen, Lei and Ni, Lionel M and others},
journal={arXiv preprint arXiv:2410.09671},
year={2024}
}

@article{o1-coder,
title={o1-coder: an o1 replication for coding},
author={Zhang, Yuxiang and Wu, Shangxi and Yang, Yuqi and Shu, Jiangming and Xiao, Jinlin and Kong, Chao and Sang, Jitao},
journal={arXiv preprint arXiv:2412.00154},
year={2024}
}

@article{latentdpl,
title={Latent inter-user difference modeling for llm personalization},
author={Qiu, Yilun and Shi, Tianhao and Zhao, Xiaoyan and Zhu, Fengbin and Zhang, Yang and Feng, Fuli},
journal={arXiv preprint arXiv:2507.20849},
year={2025}
}

@article{nextquill,
title={NextQuill: Causal Preference Modeling for Enhancing LLM Personalization},
author={Zhao, Xiaoyan and You, Juntao and Zhang, Yang and Wang, Wenjie and Cheng, Hong and Feng, Fuli and Ng, See-Kiong and Chua, Tat-Seng},
journal={arXiv preprint arXiv:2506.02368},
year={2025}
}

@article{qwen2.5,
title={Qwen2.5 Technical Report}, 
author={Qwen and An Yang and Baosong Yang and Beichen Zhang and Binyuan Hui and Bo Zheng and Bowen Yu and Chengyuan Li and Dayiheng Liu and Fei Huang and others},
year={2025},
journal={arXiv preprint arXiv:2412.15115},
}

@inproceedings{llm-trsr,
author = {Zheng, Zhi and Chao, WenShuo and Qiu, Zhaopeng and Zhu, Hengshu and Xiong, Hui},
title = {Harnessing Large Language Models for Text-Rich Sequential Recommendation},
year = {2024},
isbn = {9798400701719},
publisher = {Association for Computing Machinery},
address = {New York, NY, USA},
booktitle = {Proceedings of the ACM Web Conference 2024},
pages = {3207–3216},
numpages = {10},
location = {Singapore, Singapore},
series = {WWW '24}
}

@inproceedings{cicl,
author = {Gao, Xiang and Das, Kamalika},
title = {Customizing language model responses with contrastive in-context learning},
year = {2024},
isbn = {978-1-57735-887-9},
publisher = {AAAI Press},
booktitle = {Proceedings of the Thirty-Eighth AAAI Conference on Artificial Intelligence and Thirty-Sixth Conference on Innovative Applications of Artificial Intelligence and Fourteenth Symposium on Educational Advances in Artificial Intelligence},
articleno = {2012},
numpages = {8},
series = {AAAI'24/IAAI'24/EAAI'24}
}

@article{amazon2023,
title={Bridging Language and Items for Retrieval and Recommendation},
author={Hou, Yupeng and Li, Jiacheng and He, Zhankui and Yan, An and Chen, Xiusi and McAuley, Julian},
journal={arXiv preprint arXiv:2403.03952},
year={2024}
}


\end{document}